\documentclass[useAMS,usenatbib,epsf,psfig]{mn2e}
\usepackage{amsmath, amsfonts, amssymb,epsfig}

\begin{document}
\renewcommand{\baselinestretch}{1}

\def\msun{{\rm M}_\odot}
\newcommand{\lsimeq}{\mbox{$\, \stackrel{\scriptstyle <}{\scriptstyle
\sim}\,$}}
\newcommand{\gsimeq}{\mbox{$\, \stackrel{\scriptstyle >}{\scriptstyle
\sim}\,$}}
\title[Magnetars]{Origin and evolution of magnetars}

\author[L. Ferrario \&  D. T. Wickramasinghe]
{Lilia Ferrario and Dayal Wickramasinghe \\
Department of Mathematics, The Australian National University,
Canberra, ACT 0200, Australia}

\date{Accepted. Received ; in original form}

\def\Msun{{\rm M}_\odot}
\def\rsun{{\rm R}_\odot}
\def\lsun{{\rm L}_\odot}
\def\gradi{\ifmmode{^\circ}else$^\circ$\fi}
\def\reference{\parskip 0pt\par\noindent\hangindent 0.5 truecm}
\maketitle

\begin{abstract}
We present a population synthesis study of the observed properties of the
magnetars investigating the hypothesis that they are drawn from a population
of progenitors that are more massive than those of the normal radio pulsars.
We assume that the anomalous X-ray emission is caused by the decay of a
toroidal or tangled up field that does not partake in the spin down of the
star. Our model assumes that the magnetic flux of the neutron star is
distributed as a Gaussian in the logarithm about a mean value that is
described by a power law $\Phi=\Phi_{0}\left(\displaystyle{\frac{M_{\rm
p}}{9\Msun}}\right)^{\gamma}$~G~cm$^2$ ($8 \Msun \le M_{\rm p}\le 45\Msun$)
where $M_{\rm p}$ is the mass of the progenitor.  We find that we can explain
the observed properties of the magnetars for a model with $\Phi_{0}=2 \times
10^{25}$~G~cm$^2$ and $\gamma=5$ if we suitably parametrise the time evolution
of the anomalous X-ray luminosity as an exponentially decaying function of
time. Our modelling suggests that magnetars arise from stars in the high mass
end ($20\Msun \le M_{\rm p} \le 45\,\Msun$) of this distribution. The lower
mass progenitors are assumed to give rise to the radio pulsars.

The high value of $\gamma$ can be interpreted in one of two ways.  It may
indicate that the magnetic flux distribution on the main sequence is a strong
function of mass and that this is reflected in the magnetic fluxes of the
neutron stars that form from this mass range (the fossil field hypothesis).
The recent evidence for magnetic fluxes similar to those of the magnetars in a
high fraction ($\sim 25$\%) of massive O-type stars lends support to such a
hypothesis.  Another possibility is that the spin of the neutron star is a
strong function of the progenitor mass, and it is only for stars that are more
massive than $\sim 20\Msun$ that magnetar-type fields can be generated by the
$\alpha-\omega$ dynamo mechanism (the convective dynamo hypothesis).  In
either interpretation, it has to be assumed that all or a subset of stars in
the mass range $\sim 20\Msun - 45\Msun$, which on standard stellar evolution
models lead to black holes via the formation of a fall-back disc, must give
rise to magnetars.

Unlike with the radio pulsars, the magnetars only weakly constrain the birth
spin period, due to their rapid spin-down.  Our model predicts a birthrate of
$\sim 1.5-3\times 10^{-3}$~yr$^{-1}$ for the magnetars.

\end{abstract}


\section{Introduction}

The anomalous X-ray pulsars (AXPs) and Soft Gamma Repeaters (SGRs) have spin
down properties indicative of magnetic fields of $\sim 10^{14-15}\,$G which
places them at the high end of the neutron star field distribution.  The
source of the quiescent X-ray luminosity in both the SGRs and the AXPs is
unclear, but it is generally believed that it is associated with the decay of
some component of the magnetic field.

The standard model of \citet{Duncan92} assumes that the fields are generated
at the time of formation of the neutron star by an efficient $\alpha- \omega$
dynamo that operates at low Rossby numbers and requires millisecond birth
periods.  Neutron stars born with initial periods $P_i$ are predicted to
generate large scale magnetic fields of $3\times 10^{17}\,{\rm G}\left
({1\,\rm ms/P_i}\right)$ under optimum conditions, which is more than adequate
to explain the fields in magnetars.

A consequence of the rapid initial spin predicted for these models is that the
supernova explosions that create the magnetars are expected to be an order of
magnitude more energetic than ordinary core-collapse supernovae, if one makes
the standard assumption that angular momentum is lost by magnetic braking and
not by gravitational radiation or due to emission in a jet. However, the
energetics of some well studied supernova remnants associated with magnetars,
appear to suggest that their formation may not always be accompanied by a
hypernova explosion \citep{Vink06}. The origin of magnetic fields in magnetars
therefore remains unclear.

Recent observations have provided two additional clues on the origin of
magnetars. Firstly, several magnetars have been associated with massive star
clusters. The magnetar candidate CXO J164710.2-455216 has been associated with
the open cluster Westerlund~1 \citep{Muno06} and SGR 1806-20 with the giant
Galactic HII complex W31 \citep{Figer05}, both suggesting progenitors with
$M\gsimeq 40\Msun$. Massive ($\gsimeq 20\Msun$) progenitors appear also to be
indicated for SGR 1627-41 \citep{Corbel99} and SGR 0526-66 \citep{Klose04}.
Interestingly, the progenitors have masses in the range $\sim 20 -45\Msun$
where interaction with a fall-back disc is expected to play an important role
in determining the ultimate outcome of stellar evolution - namely whether a
neutron star or a black hole forms \citep{Heger03}.

A second clue comes from the discovery that more massive stars tend to have
higher magnetic fluxes than less massive stars, and that the incidence of
magnetism on the main sequence increases with mass.  Some $25$\% of O and
early B stars in the Orion cluster are strongly magnetic with fields in the
range $600 - 1500$~G (\citet{Petit08}, \citet{Donati02}) so that the
progenitors of the magnetars are also likely to be strongly magnetic.  Fossil
magnetic fluxes similar to those observed in the magnetars may already be
present in stellar cores prior to collapse and could give rise to global
magnetic fields of $\sim 10^{15}$~G either directly through magnetic flux
conservation, or indirectly by acting as seed fields for a dynamo. In this
context, we note that while the dynamo mechanism is effective in generating
weak fields of the order of the equi-partition value in the convective cores
of stars with radiative envelopes \citep{Brun05}, much stronger fields can be
generated by the dynamo mechanism if a fossil field is present in the
radiative region \citep{Featherstone06}.

In this paper, we carry out a population synthesis study to analyse the
observed numbers of magnetars and their magnetic field and period
distributions allowing for X-ray selection effects. Our basic premise is that
the birth magnetic field of a neutron star is related primarily to the mass of
the progenitor star $M_{\rm p}$. Our aim is to determine the form of this
relationship in the hope that it may provide some insights into the origin of
magnetars.

\section{The population synthesis model}\label{periods}
\subsection{Modelling the spin down of magnetars}\label{origin}

We assume that the magnetic field of a magnetar is composed of two components,
as in \citet{Pons07}: an underlying dipolar component $B_d$ and a crustal
component $B_c$. We assume that the dipolar component is a global field that
does not decay on a time scale of $\sim 10^8$~yr, the lower limit to the field
decay time scales in normal radio pulsars (e.g. see the most recent population
syntheses calculations of \citet{Vranesevic04}, \citet{Ferrario06} (hereafter
FW), \citet{Faucher06}) and is solely responsible for the spin down of the
magnetar which occurs along constant field lines. The crustal component is
envisioned to be a toroidal or tangled field which does not partake in the
spin down of the star, decays on time scales of $\sim 10^4-10^5$~yr, and is
the source of the anomalous X-ray and $\gamma-$ray emission (see section \ref
{results} for a discussion of the alternative hypothesis that the spin down is
due to a decaying field).

The apparent upper limit of $\sim 12$~s for the spin periods of magnetars has
to be attributed to the expected rapid decline in the X-ray luminosity
predicted by field decay models (e.g. see the period clustering study of
\citet{Colpi00}). Models of neutron stars with a heat source in the crust
predict that cooling will occur initially at a nearly constant luminosity
after which the luminosity will decay exponentially to negligible values
\citep{Kaminker06}, but there are no detailed models that investigate the
dependence of decay rates on initial magnetic field and envelope
composition. In the absence of such calculations, we have adopted the
following parametrisation to describe the decline of the crustal luminosity
$$L=L_0 \left(\frac {B_d}{10^{13}~{\rm G}}\right) 
e^{-\displaystyle{\frac{t}{\tau_d}}}$$
where 
$$\tau_d= \tau_{d0}\left(\frac{10^{13}~{\rm G}}{B_d}\right)^\delta ~~{\rm
yr}.$$ Here, $L_0$ and $\tau_{d0}$ are constants, and we have assumed that the
birth crustal luminosity is proportional to the intrinsic dipolar field. We
have also allowed for the possibility that the decay time is a function of the
magnetar's field strength $B_d$ via the parameter $\delta$.

\subsection{Parametrisation of magnetic field}\label{parametrisation}

Recently, there have been detailed stellar evolution calculations for stars in
the mass range $12 \Msun \le M_{\rm p}\le 35\Msun$ that have attempted to
incorporate magnetic fields and rotation \citep{Heger05}. These calculations
allowed for fields generated in the differentially rotating radiative regions
of stars and their effect on the transport of angular momentum, on the
assumption that a dynamo mechanism is in operation.  However, the basic
premise on which these calculations were made has recently been seriously
challenged by \citet{Zahn07}. Therefore, the role played by magnetic fields
and rotation in the evolution of stars still remains uncertain.

It is likely that the magnetic flux of the progenitor star, $\Phi_{\rm p}$,
and to a lesser extent the initial angular momentum, both play an important
role in determining the magnetic flux $\Phi_{\rm NS}$ of the nascent neutron
star.  In our modelling, we assume that the progenitor mass $M_{\rm p}$ is the
primary variable that describes the birth magnetic flux of the neutron star.

Our modelling procedure is as discussed in FW, except for our treatment of the
birth magnetic flux and birth spin period.  Briefly, we begin with a dynamical
model for the Galaxy, an initial mass function, and an initial-final mass
relationship for neutron stars. We then follow the spin evolution and motion
of neutron stars and calculate the properties of the currently
observed population of magnetars allowing for X-ray selection effects.

In the present study, we assume that the birth magnetic flux of the neutron
star is distributed as a Gaussian about a mean that is given by a power law
\begin{equation}
\label{Bf} 
\Phi_{\rm NS}=\Phi_{0} \left(\frac{M_{\rm p}}{9\Msun}\right)^{\gamma}~~ \rm{G
~cm}^2\qquad\qquad(8\Msun \le M_{\rm p}\le 45\Msun).
\end{equation}
with dispersion $\sigma_{\log\Phi_{\rm NS}}$. For consistency with our
previous study of radio pulsars, we assume $\log\Phi_0 =25.2$ and deduce the
exponent $\gamma$ from the study of magnetars.  The birth spin period is
assumed to be given by a Gaussian distribution with mean $P_0$ and dispersion
$\sigma_{P_0}$.

The birth magnetic field of the neutron stars is calculated from
\begin{equation}
\label{Bfossil} 
B_{\rm NS}=\frac{\Phi}{\pi (R_{\rm NS})^2}~~ {\rm G}
\end{equation}
where $R_{\rm NS}$ is the radius of the neutron star in cm given by the
mass-radius relationship for neutron stars which depends on the equation of
state (see FW for further details). 

\subsection{Allowing for X-ray selection in the {\it ROSAT} 
survey}\label{xray_selection}

We have chosen to use the subset of magnetars discovered from the {\it ROSAT}
All-Sky survey to constrain our models. In order to achieve this, we need to
account for interstellar extinction and X-ray selection effects.

We have implemented the methodology laid out by \citet{Gill07} for finding the
percentage of the magnetars in our calculations that yield a {\it ROSAT}
limiting count rate of 0.015 counts~s$^{-1}$ in the $0.1-2.4$~keV energy range
\citep{Hunsch99}.

Our synthetic objects have been assigned a two-component model (power law and
blackbody) modulated by interstellar absorption. Hence, as in \citet{Gill07},
the number of photons in the energy range $E$~$-$~$E+dE$ is given by

\begin{eqnarray}
 N(E)\,dE&=&\left[\eta_1\left(\frac{E}{1{\rm keV}}\right)^{-\alpha}+
   \right. \nonumber \\
& & \left.\eta_2\frac{8.1E^2}{k^4T^4\big(e^{\displaystyle{E/kT}}
-1\big)}\right]e^{\displaystyle{-N_H\sigma(E)}}\,dE
\end{eqnarray}

Here, $\eta_1$ is the power-law normalisation in units of photons
keV$^{-1}$cm$^{-2}$s$^{-1}$, $\alpha$ is the power law energy index, $kT$ is
the temperature in keV and $\eta_2$ is the blackbody normalisation in units of
$L_{39}/D_{10}^2$, where $L_{39}$ is the luminosity in units of $10^{39}$
erg~s$^{-1}$ and $D_{10}$ the distance in units of 10 kpc. The neutral
hydrogen column density $N_H$ is obtained using the \citet{Foster03} model of
the galactic HI distribution. The photoelectric absorption cross-sections
$\sigma(E)$ are from \citet{Balucinska92} used with the ISM abundances of
\citet{Wilms00}. In order to convert the absolute photon fluxes into the
photon counts of the X-Ray telescope (XRT) detector, we have used the {\it
ROSAT} calibration guidelines and the XRT effective area combined with the
PSPC sensitivity.

In our modelling, the magnetars have been assigned a temperature and an energy
index according to Gaussian distributions with means and spreads
$kT=0.5,\sigma_T=0.1$ and $\alpha=3.5,\sigma_\alpha=0.5$ respectively.  The
relative contribution of the black body and power law components to the total
X-ray luminosity has been randomly assigned with values consistent with the
observed distribution given by \citet{Perna01}.

\section{Results and discussion}\label{results}

We have computed a series of models for different values $M^*$, $\gamma$,
$\sigma_{\log\Phi0}$, $P_{0}$, $\sigma_{P0}$, $L_0$, $\tau_{d0}$ and $\delta$
until a model is obtained that satisfies the following criteria: (a) the model
must envelope and provide a reasonable representation of the observed period
and magnetic field distributions and (b) the total number of observed
magnetars must be less than the theoretically predicted number after X-ray
selection effects are taken into consideration. Given the small
number of magnetars and the unknown number of active, but currently dormant
magnetars (e.g. objects such as SGR~1627-41 that  has only recently
re-activated after a nearly 10 year long quiescent period \citep{Palmer08}),
the real Galactic distributions is not well enough sampled to justify a more
detailed modelling.

\begin{figure*}
\label{best_fit}
\begin{center}
\hspace{0.1in}
\epsfxsize=0.7\textwidth
\epsfbox[35 286 586 703]{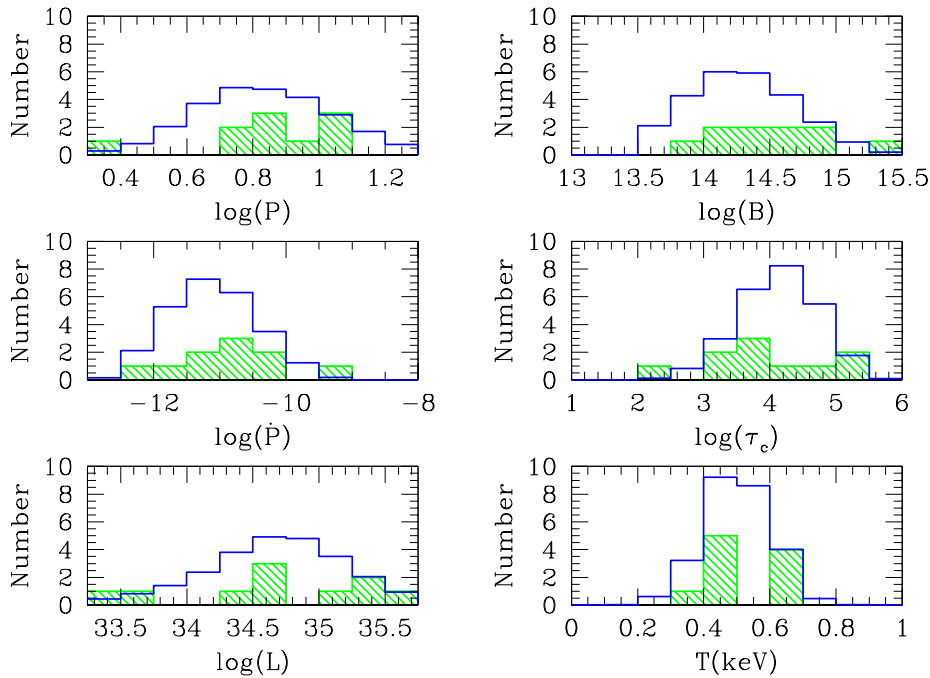}
\caption{Our model (solid line) overlapped to the magnetars data sample
  (dashed histograms) taken from
  http://www.physics.mcgill.ca/~pulsar/magnetar/main.html and with the
  addition of the recently confirmed member of the AXP class,
  1E1547.0-5408 (=PSR~J1550-5418) \citep{Camilo07}.}
\end{center}
\end{figure*}

The total number of active magnetars in the Galaxy depends strongly on $M^*$
because of the rapidly declining initial mass function. Our calculations show
that in addition to the initial mass function, the parameters $\gamma$,
$\Phi_{0}$, $\sigma_{\log\Phi_{0}}$, $L_0$, $\tau_{d0}$ and $\delta$ together
determine the magnetic field, period, and period derivative distributions.

Our results are rather insensitive to $P_{0}$ and $\sigma_{P_0}$ because of
the rapid initial spin down of magnetars. Hence, this study cannot exclude the
possibility that magnetars are born at millisecond periods as required by some
recent models \citep{Geppert06}, or that the birth spin period may be a
function of magnetic field as assumed in FW.  For simplicity, in our models we
have adopted $P_0=3$ seconds and $\sigma_{P_0}=1$ second, although the FW
prescription with an upper initial period cutoff at $8$ seconds is also
acceptable.

We present in Figure~1 our favoured model with $M^*=20\Msun$, $\log\Phi_{0} =
25.2$, $\sigma_{\log\Phi_{0}} = 0.4$, $L_0=10^{34}$, $\tau_{d0}=5\times 10^5$~yr,
and $\delta=1.3$. The corresponding birthrate is $3\times
10^{-3}$~yr$^{-1}$. This model best envelopes the field, period, period
derivative, characteristic age, luminosity and temperature distributions and
predicts 26 active magnetars after applying the {\it ROSAT} detectability
criteria.  We exclude the possibility of a significantly lower $M^*$
(e.g. $\le 18\Msun$) on the grounds that it will lead to a conflict with the
birth rate of radio pulsars which we assume arise from stars with $M\le
M^*$. On the other hand, the use of a significantly higher $M^*$ ($> 22
\Msun$) leads to too few magnetars and inconsistencies with the observed
distributions.

We have also constructed a $P-\dot {P}$ diagram which is shown in Figure 2
where our {\it ROSAT} selected objects are compared with the observations of
all magnetars with known $\dot P$. Thus, among the observed objects we have
also included SGR~0526-66, which is located in the LMC, and
CXOU~J010043-721134 which is in the SMC, since their position in the $P-\dot
{P}$ diagram is unlikely to depend on which galaxy they reside.

The predicted number of active magnetars from our favoured model can be
reconciled with the observed number of 5 {\it ROSAT} detected sources if we
assume that only $\sim 20$\% of magnetars were in an active state which
brought them to a high enough luminosity to be detected by this satellite. In
this case, the actual birthrate of magnetars would be $\sim 3\times
10^{-3}\,$yr$^{-1}$, which of course depends on our basic assumption that all
stars in the mass range $20-45\Msun$ produce magnetars. However, we note that
the total number of observed magnetars in the Galaxy is only 14. If no
additional magnetars exist in our Galaxy, that is if all magnetars have
already been discovered, this means that we have produced too many magnetars
from progenitors in the mass range $20-45\Msun$. This may suggest that some
50\% of stars in this mass range will produce black holes. If this is the
case, the birth rate of magnetars would reduce to $\sim 1.5\times
10^{-3}\,$yr$^{-1}$.

Studies of SGR bursts conducted by \citet{Kouveliotou94} indicated that there
cannot be more than 7 active SGRs in the Galaxy and \citet{Kouveliotou98}
suggested that magnetars are born at a rate of about $0.1$ per century.  The
recent study carried out by \citet{Gill07} was based on a study of the five
AXPs detected in the {\it ROSAT} All-Sky Survey, and indicated a birthrate of
0.22 per century with their progenitors being massive main sequence
stars. Both these estimates are generally consistent with the birth rate that
we deduce. Another recent study carried out by \citet{Muno08} who searched for
magnetars in archival Chandra and XMM-Newton observations of the Galactic
plane, yielded a birthrate in the range $3\times 10^{-3}-6\times 10^{-2}\,$
yr$^{-1}$. The upper end of this range is excluded by our calculations.

On the fossil field hypothesis, the high value indicated for the magnetic flux
index $\gamma$ may simply reflect the intrinsic magnetic flux distribution on
the main sequence.

Observations of main sequence stars show that the maximum magnetic fluxes
observed on the main sequence map on to the magnetic fluxes of the highest
field magnetic white dwarfs ($10^8-10^9$~G ) and neutron stars
($10^{14}-10^{15}\,$G) rather well. However, due to sensitivity limitations of
polarimetric observations, only the upper end of the magnetic field
distribution of main sequence stars (the strongly magnetic stars) can be
observed, so we only have a partial picture of magnetism on the main
sequence. Unfortunately, most stars in the mass range $8-20\Msun$ that give
rise to radio pulsars have magnetic fluxes well below the currently observed
range on the Main Sequence so that the index $\gamma$ cannot be empirically
estimated. We note however that magnetic fluxes similar to those inferred in
magnetars occur in $\sim 25$\% of massive B and O-type stars.

If we somehow dismiss the close similarities between the magnetic fluxes of
massive main sequence stars and magnetars by putting them down to mere
coincidence, alternatives to the fossil field hypothesis need to be explored.
It is possible that the mass of the progenitor determines the spin of the
nascent neutron star and thereby the strength of a dynamo generated field.
Support for this hypothesis comes from the calculations of \cite{Heger05}
which allow for angular momentum transport by magnetic fields generated by
differential rotation during stellar evolution. These show that more massive
stars tend to produce more rapidly spinning neutron stars.  For a $35\Msun$
progenitor (the highest mass they considered), they predict that the neutron
star will have a spin period of $3$~ms which is rapid enough for the
generation of a a magnetar-type field by the $\alpha-\omega$ dynamo mechanism
of \citet{Duncan92}. However, for this to be a viable explanation of the
magnetars, the $\alpha-\omega$ dynamo mechanism would need to be effective for
neutron stars that arise from progenitors with significantly lower masses
($\sim 20-22\Msun$) or the derived birthrate of magnetars would be too low to
explain the observations. Furthermore, since the calculations of
\citet{Heger05} have been seriously challenged by \citet{Zahn07}, it appears
that we are still waiting for more stellar evolution calculations, that allow
for both magnetic field and rotation, to show us whether or not millisecond
rotation periods can result from massive stars. If such periods can be
achieved, this would give support to the idea that the dynamo model is a
viable alternative to the fossil field hypothesis, although it is likely that
a fossil field could still play the role of a seed field.

\begin{figure}
\begin{center}
\epsfxsize=0.4\textwidth
\epsfbox[32 168 524 646]{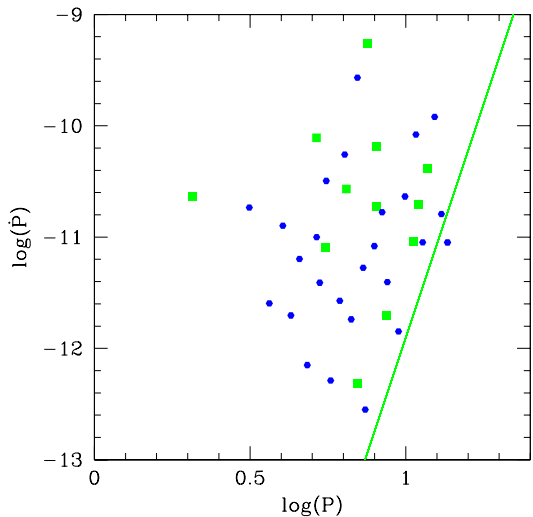}
\caption{Filled squares: observed magnetars
(http://www.physics.mcgill.ca/~pulsar/magnetar/main.html) with the addition of
1E1547.0-5408 \citep{Camilo07}; filled circles: magnetars as derived from our
modelling (see text). The solid line is an empirically determined boundary (a
``magnetar death line'') given by $\log(\dot P)=8.4\log(P)-20$. }
\end{center}
\label{ppdot_diag}
\end{figure}

It is relevant that we comment on the alternative to our basic hypothesis on
spin evolution, namely that the field that decays is also the field that
drives spin evolution during the magnetar phase. We have considered this
possibility using ``Avenue C'' of \citet{Colpi00}, but excluded it from our
present considerations because it failed to populate the high field end of the
observed magnetar field distribution. It is conceivable that with the use of
different field decay parameters for avenue C we may also be able to model the
data, but we expect that such a model will need to have field characteristics
during the magnetar phase that are so similar to those adopted in the present
calculations that our major conclusions will remain largely unchanged.

In conclusion, we note that the origin of the crustal field component in
magnetars remains unresolved.  \citet{Zahn07} have shown that the shearing of
a poloidal field of fossil origin in a differentially rotating radiative
region can lead to the generation of a toroidal field so that field complexity
of the type that appears to be required in magnetars may arise naturally from
stellar evolution, when the complex interplay between rotation and fossil
magnetic fields is taken into consideration. On the other hand, since
fall-back discs are expected to play a role in the evolution of stars in the
mass range $\sim 20 - 45\Msun$, it is tempting to speculate that the crustal
field component that characterises the magnetars may have its origin in the
interaction of the fall-back disc with the magnetic field of the stellar core,
in cases where a magnetar is the outcome of stellar evolution. If this is the
case, one would expect that most of the mass in the fall-back disc would be
magnetically ejected.

\section*{Acknowledgements}

The authors would like to thank the Referee, U. Geppert, for valuable comments
and Koji Mukai for providing assistance with the ROSAT/PSPC X-ray selection
software.

\end{document}